\documentclass[runningheads,a4paper]{llncs}
\usepackage{amssymb,amsmath,latexsym}
\usepackage{graphicx}
\usepackage{verbatim}
\usepackage{xcolor}
\usepackage{algorithm,algorithmic}
\usepackage{float} % To force a float to remain in a specific location, add the float package to your preamble and use the [H] float placement specifier. Done.
\definecolor{red}{RGB}{255, 0, 0}
\definecolor{blue}{RGB}{0, 0, 255}

\usepackage{url}
\urldef{\mailsa}\path|{mghasemineja2017, mnojoumian}@fau.edu|
\newcommand{\keywords}[1]{\par\addvspace\baselineskip
\noindent\keywordname\enspace\ignorespaces#1}

\begin{document}

\mainmatter  % start of an individual contribution

\title{Secure Error Correction using Multi-Party Computation}

\titlerunning{Secure Error Correction using Multi-Party Computation}

\author{Mohammad G. Raeini \and Mehrdad Nojoumian}
\authorrunning{Secure Error Correction using Multi-Party Computation}

\institute{Department of Computer and Electrical Engineering and Computer Science\\Florida Atlantic University, Boca Raton, FL, USA\\
\mailsa\\}


\toctitle{Lecture Notes in Computer Science}
\tocauthor{Authors' Instructions}
\maketitle

\begin{abstract}
During recent years with the increase of data and data analysis needs, privacy preserving data analysis methods have become of great importance. Researchers have proposed different methods for this purpose. Secure multi-party computation is one of such techniques that allows a group of parties to evaluate a function on their data without revealing the data. This is done by secret sharing approach, in which parties share a piece of their data using polynomials and after doing function evaluation on shares of data finally they do a Lagrange interpolation to get the result. Two approaches have been proposed in secure multi-party computation for evaluating a function, arithmetic gates and logical gates. In both of them and since communication is an important step in multi-party computation, errors may happen. So, being able to detect and correct errors is important. Moreover, as adversaries may interrupt communication or manipulate the data, either in communication or during computation, this error detection and correction provide participating parties with a technique to detect such errors. Hence, in this paper we present a secure multi-party computation error correcting technique that has the ability to detect and correct errors on players shares. This technique is based on Berlekamp-Welch error correcting codes and we assume that players shares are generated using Reed-Solomon codes.
\keywords{multi-party computation, error correcting codes, Reed-Solomon codes, Berlekamp-Welch algorithm}
\end{abstract}

\section{Introduction}
During recent years, the increase of data on one hand and the importance of privacy of them on the other hand, encouraged researchers to work on techniques that allow data holders to do data analysis techniques without revealing private data. One of such techniques is secure multi-party computation.
Secure multi-party computation is a kind of distributed computation that allow a set of parties to compute a function on their private data, without revealing them. The idea was first introduced by Yao, in his Millionarie's problem \cite{AYao1982}. The problem says two millionaries want to know who is richer, but they do not want to reveal their wealth.
After Yao's seminal paper, many researches have been done in this direction and his idea was generalized to a general purpose form. 
Specifically, researchers focused on how they can evaluate a function on \textit{n} piece of data, holding by \textit{n} parties, without revealing the data. It has been shown that any function can be computed in on such data without revealing them. Two general approaches have been proposed, using binary circuits or using arithmetic circuits. In the first case, \textit{n} parties share their data among themselves, using for example Shamir's secret sharing \cite{AdiShamir1979} and then the parties do function calculation using binary circuits on the shared data. In this approach, parties do bit-wise calculation using for example Oblivious Transfer. As communication is an inseparable part of multi-party computation, because parties are constantly send and receive data, this approach is not efficient from communication complexity point of view. In the arithmetic circuit scenario, the parties use secret sharing to share their data and do function evaluation on them. As this idea is based on polynomials and finite field calculation, it is more efficient.\\
In both scenarios all parties need to communicate and share their data. On the other hand adversaries may interrupt data communication, or create noise while players are doing local calculation on their data. As such, if they have the ability to use an error detection and correction technique, they can increase the reliability of the protocol they use for multi-party computation. In this paper we propose a secure error correction technique that can be added to a secure multi-party computation for detecting errors created by adversaries or communication channels. The idea is based on Berlekamp-Welch algorithm and we assume parties have created their shares based on Reed-Solomon codes. The reason that we use these two specific techniques is that, as secret sharing, they are based on polynomials over finite fields which are efficient for implementation.\\
The paper organization is as follows. First we presents the preliminaries of our work, including multi-party computation, Reed-Solomon codes and their decoding using Berlekamp-Welch algorithm. Then the previous works will be reviewed. After that the main part of this paper will be presented.
%%%%%%%%%%%%%%%%%%%%%%%%%%%%%%%%%%%%%%%%%%%%%%%%%%%%%%%%%%%%%%%%%%%%%%%%%%%%%%%%%%%

\section{Preliminaries}
In this section we present the preliminaries our our work. These include multi-party comutation, Reed-Solomon codes (we call them RS codes for simplicity), Berlekamp-Welch decoding algorithm (we call it BW algorithm for simplicity) for RS codes.

%%%%%%%%%%%%%%%%%%%%%%%%%%%%%%%%%%%%%%%%%%%%%%%%%%%%%%%%%%%%%%%%%%%%%%%%%%%%%%%%%%%
\subsection{Secure Multi Party Computation}
Secure multi-party computation which its first idea was introduced by Yao \cite{AYao1982} is defined as follows. $n$ parties each have a private value, want to evaluate a public function on their private data in such a way that they don't reveal any information about their data but they all can get the output of the public function. One simple example is that $n$ parties have $n$ integer values and they want to find the maximum of them without revealing them.\\
Many researches have been done in this area \cite{T_Nishide_2007}, \cite{DamgardFitziEtAll2006}, \cite{ChaumCrepeauDamgard1988}, \cite{BenorGoldwasseWigderson1988}, \cite{GoldreichMicaliWigderson1987}. In secure multi-party computation two approaches can be used. One approach is doing computation based on bits of shared values and the other approach is doing computation based on shared values of secrets in a finite field $Z_p$, for a prime integer p, \cite{T_Nishide_2007}. In both cases the shares of secret values are being distributed using a secret sharing scheme, such as Shamir secret sharing \cite{AdiShamir1979}. However, both approaches have their own pros and cons, for example,
doing addition and multiplication is efficient in finite field arithmetic, but using boolean circuits it is not efficient. Unfortunately, doing calculations such as secret comparisons is not efficient and trivial in arithmetic circuits, whereas it is trivial in boolean circuit calculations \cite{T_Nishide_2007}. However, for large integers this task is not efficient using boolean circuits.\\
To overcome the inefficiency of these two scenarios and having an efficient in \cite{DamgardFitziEtAll2006} authors presented a protocol, called bit-decomposition, that allows parties to convert sharing of finite field elements to sharing of bits. In \cite{T_Nishide_2007} authors improved the bit-decomposition protocol by reducing its communication complexity. Another work in this area has been presented in \cite{SchoenmakersTuyls2006} which is based on threshold homomorphic systems.

%%%%%%%%%%%%%%%%%%%%%%%%%%%%%%%%%%%%%%%%%%%%%%%%%%%%%%%%%%%%%%%%%%%%%%%%%%%%%%%%%%%
\subsection{Reed-Solomon Codes}
Reed-Solomon codes was introduced in 1960 in \cite{ReedSolomon1960}. These linear error-correcting codes are based on polynomials over finite fields and have many applications. In the following we assume all calculation are done in the finite field $Z_p$ for a given prime number $p$. RS codes encode a message of length $k$ into a codeword of length $n$, $k \leq n \leq p$. Mathematically, given a message $m = [m_0, m_1, m_2, ..._, m_{k-1}]$, the polynomial $P$ is defined as following:
\begin{equation} \label{RS_Code_poly_eq1}
P(x) = m_0 + m_1 x + m_2 x^2 + ... + m_{k-1} x^{k-1}
\end{equation}
In which coefficients are in $Z_p$. To encode the message $m$, the polynomial will be evaluated on $n$ different points, say $1, 2, ..., n$, so the encoded message, denoted by $c$ would be:

\begin{equation}
c = [c_0, c_1, c_2, ..._, c_{n-1}] = [P(1), P(2), P(3), ..., P(n)]
\end{equation}
For the Reed-Solomon codes we have the following theorems:
\begin{theorem}
The weight of Reed-Solomon codes of length $n$ with a message of length $k$, $RS(n, k)$, is $n - k + 1$.
\end{theorem}

\begin{theorem} \label{RS_Errors_theorem}
An $RS(n, k)$ with $n = k + 2e$ can correct $e$ errors.
\end{theorem}

A decoding algorithm was developed by Berlekamp and Welch in \cite{BerlekampWelch1986}, which we will discuss it in the next section.

%%%%%%%%%%%%%%%%%%%%%%%%%%%%%%%%%%%%%%%%%%%%%%%%%%%%%%%%%%%%%%%%%%%%%%%%%%%%%%%%%%%
\subsection{Berlekamp-Welch Decoding Algorithm}
Berlekamp-Welch algorithm is a decoding algorithm for RS codes \cite{BerlekampWelch1986}. As we discussed in the previous section, an $RS(n, k)$ code encodes a message with length $k$ to a codeword of length $n$. Now, assume that $e$ errors has happened in the codeword
$c = [c_0, c_1, c_2, ..._, c_{n-1}] = [P(1), P(2), P(3), ..., P(n)]$
and the codeword with error is 
$r = [r_0, r_1, r_2, ..._, r_{n-1}]$.
That is, $r_i \neq P(i)$ for at most $e$ cases.
\begin{theorem} \label{theorem: bw alg theorem}
Given a received codeword, generated by RS(n, k) codes, with $e$ errors, then there exists non-zero polynomials $E(x)$ and $Q(x)$
for which we have \cite{mHaiman_LectureNote}:
\begin{equation}
degree(E(x)) \leq e
\end{equation}
\begin{equation}
degree(Q(x)) \leq m + e - 1
\end{equation}
\begin{equation} \label{key_equation_eq}
Q(i) = r_i E(i) \hspace{1cm}  \forall i = 1, 2, ..., n.
\end{equation}
Moreover, $\frac{Q(x)}{E(x)}$ will give the polynomial that has generated the codeword, $P(x)$.
\end{theorem}
Equation \ref{key_equation_eq} is called key equation and solving it gives us the location of the errors in the received codeword, which is guaranteed by theorem \ref{theorem: bw alg theorem}.\\
For $i = 1, 2, ..., n$, key equation, equation \ref{key_equation_eq}, will produce a linear system of equations, which we can solve it by different methods in linear algebra, such as Gaussian elimination or Cramer's rule. By finding the solution of key equation, we can find errors location and the polynomial $P$ which gives the corrected message.

%%%%%%%%%%%%%%%%%%%%%%%%%%%%%%%%%%%%%%%%%%%%%%%%%%%%%%%%%%%%%%%%%%%%%%%%%%%%%%%%%%%
\section{Secure Error Detection and Correction using MPC}
We assume that $n$ parties have $n$ shares and they want to be able to check if any errors has happened in their data, because in multy-party computation parties constantly exchange their data. We also assume that all the following calculations are done in finite field $Z_p$ where $p$ is a prime number.\\
In order to be able to detect and correct $e$ errors, we need to have at least $3e + 1$ shares, in other words $3e + 1$ parties need to participate, because according to theorem \ref{RS_Errors_theorem}, $n \geq k + 2 e$, which $k$ is the message length. Also, we assume that the number of errors at most can be one less than the message length. In other words, all the message has not been altered.
%In the following we will address the problem of error detection in two scenarios. First, when one error has happened and then when more than one errors  have happened.\\
In the following we will address the problem of error detection and then provide a technique that allows the parties to recover the incorrect share.\\
%%%%%%%%%%%%%%%%%%%%%%%%%%%%%%%%%%%%%%%%%%%%%%%%%%%%%%%%%%%%%%%%%%%%%%%%%%%%%%%%%%%%%%%%%%%%%%%%%%%%%%%%%%%%%%%%%
\\
\textbf{\textit{Locating one error at the time:}}
Each player creates an equation using his secret value (which is denoted by $\alpha_i$ for player $i$).\\
%\textcolor{red}{this $a_{n-3} x^{n-3}$ was added:}\\
\begin{equation} \label{eq1}
a_0 + a_1 x + a_2 x^2 + ... + a_{n-3} x^{n-3} + a_{n-2} x^{n-2} = \alpha_i ( x + b_0)
\end{equation}
So we have $n$ equations, each at hand of one party, by which we can define the following system linear equation (consisting of $n$ equations and $n$ unknowns including $a_0, a_1, a_2, ..., a_{n-2}, b_0$).
\begin{equation}
\left\{ 
\begin{array}{c}
a_0 + a_1 x + a_2 x^2 + ... + a_{n-3} x^{n-3} + a_{n-2} x^{n-2} = \alpha_1 ( x + b_0) \\ 
a_0 + a_1 x + a_2 x^2 + ... + a_{n-3} x^{n-3} + a_{n-2} x^{n-2} = \alpha_2 ( x + b_0) \\ 
...\\
a_0 + a_1 x + a_2 x^2 + ... + a_{n-3} x^{n-3} + a_{n-2} x^{n-2} = \alpha_n ( x + b_0) \\ 
\end{array}
\right. 
\end{equation}
$a_0, a_1, a_2, ..., a_{n-2}$ will be used for the error correction polynomial $Q$ in BW algorithm and $b_0$ is error locator as the $E$ polynomial in BW algorithm. Also we assume that all of calculations are done in $Z_p$ for a public pre-defined prime number $p$.
Now, players evaluate equations with $x = 1, 2, 3, ..., n$, like BW algorithm. So, they have:
\begin{equation}
\left\{ 
\begin{array}{c}
a_0 + a_1 + a_2 + ... +  a_{n-3} + a_{n-2} = \alpha_1 ( 1 + b_0) \\ 
a_0 + 2 a_1 + 4 a_2 + ... +  2^{n-3} a_{n-3} + 2^{n-2} a_{n-2} = \alpha_2 ( 2 + b_0) \\ 
...\\
a_0 + n a_1 + n^2 a_2 + ... + n^{n-3} a_{n-3} + n^{n-2} a_{n-2} = \alpha_n ( n + b_0) \\ 
\end{array}
\right. 
\end{equation}
To make it simple, we use matrix notation to show this linear system of equations:
\begin{equation}
  \begin{bmatrix}
    1 & 1 & 1 & \hdots & 1^{n-2} & -\alpha_1 \\
    1 & 2 & 4 & \hdots & 2^{n-2} & -\alpha_2 \\
    1 & 3 & 9 & \hdots & 3^{n-2} & -\alpha_3 \\
    \vdots & \vdots & \vdots & \ddots & \vdots & \vdots  \\
    1 & n-1 & {(n-1)}^2 & \hdots & (n-1)^{n-2} & -\alpha_{n-1} \\
    1 & n & n^2 & \hdots & n^{n-2}  & -\alpha_n
  \end{bmatrix}
  \begin{bmatrix}
    a_0\\
	a_1\\
	a_2\\
	\vdots \\
	a_{n-2}\\
	b_0
  \end{bmatrix}
  =
    \begin{bmatrix}
    \alpha_1 \\
    2 \alpha_2 \\
    3 \alpha_3 \\
	\vdots \\
    (n-1)\alpha_{n-1}\\
    n \alpha_n
  \end{bmatrix}
\end{equation}
The first $n-2$ columns of the first matrix are public values, which are the same as the columns of the Vandermone matrix. If we use Cramer's rule for solving this system of equations (just for $b_0$, which determines the location of the error), we would have:
\begin{equation} \label{b0_eq}
	b_0 = \frac{det(A_1)}{det(A_2)}
\end{equation}
where
\begin{equation}
A_1 = 
\begin{bmatrix}
    1 & 1 & 1 & \hdots & 1^{n-2} & \alpha_1 \\
    1 & 2 & 4 & \hdots & 2^{n-2} & 2\alpha_2 \\
    1 & 3 & 9 & \hdots & 3^{n-2} & 3\alpha_3 \\
    \vdots & \vdots & \vdots & \ddots & \vdots & \vdots \\
    1 & n-1 & {(n-1)}^2 & \hdots & (n-1)^{n-2} & (n-1) \alpha_{n-1} \\
    1 & n & n^2 & \hdots & n^{n-2} & n \alpha_n
\end{bmatrix}
\end{equation}
and
\begin{equation}
A_2 = 
\begin{bmatrix}
    1 & 1 & 1 & \hdots & 1^{n-2} &-\alpha_1 \\
    1 & 2 & 4 & \hdots & 2^{n-2} &-\alpha_2 \\
    1 & 3 & 9 & \hdots & 3^{n-2} &-\alpha_3 \\
    \vdots & \vdots & \vdots & \ddots & \vdots & \vdots \\
    1 & n-1 & {(n-1)}^2 & \hdots & (n-1)^{n-2} & - \alpha_{n-1} \\
    1 & n & n^2 & \hdots & n^{n-2} & -\alpha_n
\end{bmatrix}
\end{equation}
if we expand the determinant based on the last column, we will have:
\begin{equation} \label{d1_eq}
d_1 = det(A_1) = \sum_{i=1}^{n} (-1)^{i + n} (-\alpha_i) det(A_1^{i, n})
\end{equation}
and
\begin{equation} \label{d2_eq}
d_2 = det(A_2) = 
\sum_{i=1}^{n} (-1)^{i + n} (\alpha_i) det(A_2^{i, n})
\end{equation}
where $A^{i, n}$ is the $(n - 1)\times(n - 1)$ matrix that is created by eliminating the $i$-th row and $n$-th column of A.
As we see in the above equation just $\alpha_i$'s are secret and the second term in the summation is a public value. So, to calculate $d_1$ and $d_2$, players just need to locally multiply their secret value by a public term and send a shares of the result to other players and finally do a Lagrange interpolation to find the location of the error. The error detection algorithm is as following, see algorithm \ref{alg:Error detection protocol}.\\
%%%%%%%%%%%%%%%%%%%%%%%%%%%%%%%%%%%%%%%%%%%%%%%%%%%%%%%%%%%%%%%%%%%%%%%%%%%%%%%%%%%%%%%%%%%%%%
\begin{algorithm}[H]% Use "stay right HERE" already!
\caption{Error Detection Protocol}
\label{alg:Error detection protocol}
\begin{algorithmic}[1]
  \STATE Each player defines his own equation (with his public ID $i$ and his secret value $\alpha_i$), as following:
\begin{equation}
Q(i) = \alpha_i E(i)
\end{equation}
\STATE All players put their public part of their equation in a matrix (in the last column which is private value of each player they just put $*$, because in the next step during Gaussian expansion, this will be eliminated) as follows:
\begin{equation}
A = 
  \begin{bmatrix}
    1 & 1 & 1 & \hdots & 1^{n-2} & * \\
    1 & 2 & 4 & \hdots & 2^{n-2} & * \\
    1 & 3 & 9 & \hdots & 3^{n-2} & * \\
    \vdots & \vdots & \vdots & \ddots & \vdots & \vdots \\
    1 & n-1 & {(n-1)}^2 & \hdots & (n-1)^{n-2} & * \\
    1 & n & n^2 & \hdots & n^{n-2}  & *
  \end{bmatrix}
\end{equation}
\STATE One of the players accepts to calculate $det(A_1^{i, n})$ and $det(A_2^{i, n})$, from $A$ matrix. $A^{i, n}$ means the $(n-1)$ by $(b-1)$  matrix that has been created from $A$ by eliminating its $i$-th row and $n$-th column. After calculation this player hands out 
$det(A_1^{i, n})$ and $det(A_2^{i, n})$ to player with ID $i$.
\\
\STATE Now, each player, which just received his related terms in equations \ref{d1_eq} and \ref{d2_eq}, calculate the multiplication of his private value by the received term locally, we call the result value $det_i$ for player $i$.
\\
\STATE Each player, shares his $det_i$ between all players.
\\
\STATE Each player adds up his received shares, lets denote the result by $s_i$, that is $s_i = \sum_{j=1}^{n} {det}_j$.
\\
\STATE Finally all players do a Lagrange interpolation on their $s_i$ and get the $d_1$ and $d_2$, as in equation \ref{b0_eq}.
Note that for $d_1$ and $d_2$ players need to multiply their private values with the corresponding terms as in equations \ref{d1_eq} and \ref{d2_eq}.
\end{algorithmic}
\end{algorithm}
%%%%%%%%%%%%%%%%%%%%%%%%%%%%%%%%%%%%%%%%%%%%%%%%%%%%%%%%%%%%%%%%%%%%%%%%%%%%%%%%%%%%%%%%%%%%%%%%%%%%%%%%%%%%%%%%%%%%%%%%%%%%%%%%%%%
%\newpage
\textbf{\textit{\\Correcting one error at a time:}}
For the error correction we will use the idea in \cite{NojoumianStinsonGrainger2010}. In this paper authors proposed a new approach, based on Lagrange interpolation, for recovering incorrect shares. After the location of error has been determined, solving the system of equations for $b_0$, the other parties who have correct shares, help the party with incorrect share to correct his share. The whole idea is that each player calculates his Lagrange interpolation constant and multiplies it by his share and send the shares of the result value to all players. After all players do so, they will have portions of a share of the original data, by which the player with incorrect share can correct his share. The error correction protocol  \ref{alg:Error correction protocol} shows the step by step process for helping a player to correct his share.
%%%%%%%%%%%%%%%%%%%%%%%%%%%%%%%%%%%%%%%%%%%%%%%%%%%%%%%%%%%%%%%%%%%%%%%%%%%%%%%%%%%
%%%%%%%%%%%%%%%%%%%%%%%%%%%%%%%%%%%%%%%%%%%%%%%%%%%%%%%%%%%%%%%%%%%%%%%%%%%%%%%%%%%%%%%%%%%%%%%%
\begin{algorithm}[H]% Use "stay right HERE" already!
\caption{Error Correction Protocol}
\label{alg:Error correction protocol}
\begin{algorithmic}[1]
  \STATE Each party $i$ calculates his Lagrange interpolation constant as following:
\begin{equation}
\gamma_i = \prod_{1 \leq j \leq t, i \neq j} (\frac{k - j}{i - j})
\end{equation}
\STATE Then each player multiplies his secret value $\alpha_i$ by his Lagrange interpolation constant $\gamma_i$ and splits it into $t$ portions, where $t$ is threshold in secret sharing.
\begin{equation}
\alpha_i \times \gamma_i = \delta_{1i} + \delta_{2i} + \dots \delta_{ti}.
\end{equation}
Next, he sends each portion to one of $t$ players.
\STATE Each party $j$ receives $t$ portions in total and adds them up as follows:
\begin{equation}
\sigma_j = \sum_{i = 1}^{t} \delta_{ji}.
\end{equation}
which $\delta_{ji}$ means the portion that participant $i$ has sent to participant $j$.
\STATE Party $j$ then sends $\sigma_j$ to party, say with ID $k$, whose share is corrupted or altered. $t$ players need to send their share for player $k$, so he will be able to recover his correct share.
\STATE Party $k$ adds up all the $t$ received shares, from helper players, the result is his recovered share:
\begin{equation}
\alpha_k^{corrected} = \sum_{j = 1}^{t} \sigma_{j}.
\end{equation}
\end{algorithmic}
\end{algorithm}
%%%%%%%%%%%%%%%%%%%%%%%%%%%%%%%%%%%%%%%%%%%%%%%%%%%%%%%%%%%%%%%%%%%%%%%%%%%%%%%%%%%
\newpage
\section{Technical Discussion}
In this section we will discuss the security analysis as well as computational and communication complexity of the error detection and correction protocols.
\subsection{Security Analysis of Error Dectection Protocol}
Similar to $(t, n)$-threshold secret sharing, in which a group of $t$ players need to cooperate in order to get the secret value, here in this secure error correction protocol $t$ players need to come together to detect and correct an error. To do so, they create a asystem of linear equations and when they represent it in the matrix format, $A x = b$, one column of the $A$ matrix and the $b$ vector consist of secret values of players. As they calculate the determinant of the matrix using Gaussian expansion based on the column containing the secret values, the new sub-matrices are all containing public values, which their determinant will be calculated and distributed among all players by a volunteer player. Then each player needs to caclulate a multiplication by a public value localy which can be calculated easily by each player without revealing any information, , see equations \ref{d1_eq} and \ref{d2_eq} and the toy example in appendix A. 
\subsection{Round Complexity Discussion}
Communication or round complexity, which can very large in some MPC protocol, for our secure error correction protocol is very small. For error detection players calculate two determinants ($d_1$ according to the equation \ref{d1_eq} and $d_2$ according to the equation \ref{d2_eq}) in which they only do multiplication by some public values, which can be easily done and does not need any multiplication in MPC which requrires sharing and resharing. After that they do a Lagrange interpolation collaboratively.\\
For error correction, one round of communication is needed in step 2 and one round in step 3, in totall two round of communication.
\subsection{Computational Complexity Discussion}
The computational complexity of error correction algorithm is $O(n^3)$ where $n$ is the number of players. Because a volunteer player calculates the determinant of $(n - 1)$ by $(n - 1)$ matrices which can be done in $O(n^3)$, or by using more efficient algorithms in a lower time complexity. The Lagrange interpolation can be done in $O(n^2)$. 
%%%%%%%%%%%%%%%%%%%%%%%%%%%%%%%%%%%%%%%%%%%%%%%%%%%%%%%%%%%%%%%%%%%%%%%%%%%%%%%%%%%
\section{Conclusion}
In this paper we introduced the idea of secure error correction that allows a group of parties to detect and correct their shares in a privacy-preserving manner. The idea is based on Reed-Solomon codes and Berlekamp-Welch decoding algorithm. In fact we used Berlekamp-Welch algorithm for finding the location of the errors and for error correction we used the assumption of Shamir's $(t, n)$-threshold Secret Sharing, that a group of $t$ parties have enough data to recover the secret or equivalently create the share of a specific party. The protocol works as follows.\\
At any point in an MPC protocol, players can collaboratively form the system of equations in BW algorithm and then solve it. Note that each equation is at the hand of one player. Then they expand the determinant of the matrix of their system of equations using the Laplace expansion and they calculate and announce the determinant of the sub-matrices if there is no secret value in any column. After doing calculation and finding the location of the error, which is the ID of a player, any subset of $t$ of players with correct share can help the player with incorrect share to recover his correct share ($t$ is threshold in the Shamir's (t, n)-threshold secret sharing).

\newpage
\section*{Appendix: A Toy Example}
\textbf{\textit{A toy example:}}
Assume that the prime $p = 7$ and players ID are 1, 2, 3, 4. Also the private values of four parties are 2, 0, 5, 3 and the third value has changed to 4 because of an error. So, the codeword is
$[c_0, c_1, c_2, c_3] = [2, 0, 5, 3]$ and the received vector, as described in BW algorithm, is
$[r_0, r_1, r_2, r_3] = [\alpha_0, \alpha_1, \alpha_2, \alpha_3] = [2, 0, 4, 3]$.
The key equation of BW algorithm would be
\begin{equation} \label{eq:example}
a_0 + a_1 x + a_2 x^2 = \textcolor{red}{\alpha_i} (x + b_0)
\end{equation}
Evaluating it for $x = 1, 2, 3, 4$ (here each $x$ is the ID of a player and each equation is at the hand of one player):
\begin{equation}
\left\{ 
\begin{array}{c}
a_0 + a_1 + a_2 =     \textcolor{red}{2} (1 + b_0) \\
a_0 + 2 a_1 + 4 a_2 = \textcolor{red}{0} (2 + b_0) \\
a_0 + 3 a_1 + 2 a_2 = \textcolor{red}{4} (3 + b_0) \\
a_0 + 4 a_1 + 2 a_2 = \textcolor{red}{3} (4 + b_0) \\ 
\end{array}
\right. 
\end{equation}
In fact each player create his equation using his ID which is a public value.
This system of equations can be written as, after doing all calculation in $Z_7$:
\begin{equation}
\left\{ 
\begin{array}{c}
a_0 + a_1 + a_2 +     \textcolor{red}{5} e_0= 2 \\
a_0 + 2 a_1 + 4 a_2 + \textcolor{red}{0} e_0 = 0 \\
a_0 + 3 a_1 + 2 a_2 + \textcolor{red}{3} e_0 = 5\\
a_0 + 4 a_1 + 2 a_2 + \textcolor{red}{4} e_0 = 5 \\ 
\end{array}
\right. 
\end{equation}
To find the erroneous private value (or its location) we need to find $e_0$ as following, which was explained in section 3.
\begin{equation}
	b_0 = \frac{det(A_1)}{det(A_2)}
\end{equation}
where
\begin{equation}
A_1 = 
\begin{bmatrix}
    1 & 1 & 1 & \textcolor{red}{2} \\
    1 & 2 & 4 & \textcolor{red}{0} \\
    1 & 3 & 2 & \textcolor{red}{5} \\
    1 & 4 & 2 & \textcolor{red}{5} \\
\end{bmatrix}
\end{equation}
and
\begin{equation}
A_2 = 
\begin{bmatrix}
    1 & 1 & 1 & \textcolor{red}{5} \\
    1 & 2 & 4 & \textcolor{red}{0} \\
    1 & 3 & 2 & \textcolor{red}{3} \\
    1 & 4 & 2 & \textcolor{red}{4} \\
\end{bmatrix}
\end{equation}
After doing the calculation we will have $b_0 = 4$, which means the location of the error is 3. Because error locator polynomial would be
$ x + b_0 = x + 4$ and its root shows the location of the error, and its root in $Z_7$ is $3$. Notice that the last column of $A_1$ and $A_2$ are consisted of a multiplicative factor of parties private value, so the determinant calculation should be expanded based on that column and after the expansion players can use any method to calculate the determinant of the resulting $3 \times 3$ matrix.

\newpage
%%%%%%%%%%%%%%%%%%%%%%%%%%%%%%%%%%%%%%%%%%%%%%%%%%%%%%%%%%%%%%%%%%%%%%%%%%%%%%%%%%%

\end{document}